\documentclass[jmp,showpacs,showkeys,amsmath,amssymb,superscriptaddress,preprint,floatfix]{revtex4-1}

\begin{document}

\title[L\'{e}vy stable distributions via associated integral transform]
{L\'{e}vy stable distributions via associated integral transform}

\author{K. G\'{o}rska}
\email{kasia\_gorska@o2.pl}

\affiliation{H. Niewodnicza\'nski Institute of Nuclear Physics, Polish Academy of Sciences,\\
ul. Eliasza-Radzikowskiego 152,  PL 31342 Krak\'ow, Poland}

\affiliation{Instituto de F\'{\i}sica, Universidade de S\~{a}o Paulo, \\
P.O. Box 66318, B 05315-970 S\~{a}o Paulo, SP, Brasil}

\affiliation{Laboratoire de Physique Th\'eorique de la Mati\`{e}re Condens\'{e}e (LPTMC),\\
Universit\'e Pierre et Marie Curie, CNRS UMR 7600\\
Tour 13 - 5i\`{e}me \'et., B.C. 121, 4 pl. Jussieu, F 75252 Paris Cedex 05, France}

\author{K. A. Penson}
\email{penson@lptl.jussieu.fr}

\affiliation{Laboratoire de Physique Th\'eorique de la Mati\`{e}re Condens\'{e}e (LPTMC),\\
Universit\'e Pierre et Marie Curie, CNRS UMR 7600\\
Tour 13 - 5i\`{e}me \'et., B.C. 121, 4 pl. Jussieu, F 75252 Paris Cedex 05, France}


\begin{abstract}
We present a method of generation of exact and explicit forms of one-sided, heavy-tailed L\'{e}vy stable probability distributions $g_{\alpha}(x)$, $0 \leq x < \infty$, $0 < \alpha < 1$. We demonstrate that the knowledge of one such a distribution $g_{\alpha}(x)$ suffices to obtain exactly $g_{\alpha^{p}}(x)$, $p=2, 3, \ldots\,\,$. Similarly, from known $g_{\alpha}(x)$ and $g_{\beta}(x)$, $0 < \alpha, \beta < 1$, we obtain $g_{\alpha \beta}(x)$. The method is based on the construction of the integral operator, called L\'{e}vy transform, which implements the above operations. For $\alpha$ rational, $\alpha = l/k$ with $l < k$, we reproduce in this manner many of the recently obtained exact results for $g_{l/k}(x)$. This approach can be also recast as an application of the Efros theorem for generalized Laplace convolutions. It relies solely on efficient definite integration. 
\end{abstract}

\maketitle

\section{Introduction}

The one-dimensional L\'{e}vy stable one-sided distributions $g_{\alpha}(x)$, $0 < \alpha < 1$ are normalized probability density functions (PDF) defined on $0 \leq x < \infty$, with the peculiar property that their mean, variance and all higher moments diverge. The index $\alpha$ enters the principal characterization of $g_{\alpha}(x)$ via their Laplace transform
\begin{equation}\label{eq1}
\mathcal{L}\left[g_{\alpha}(x); p\right] = \int_{0}^{\infty} e^{-p\, x}\, g_{\alpha}(x)\, dx = e^{- p^{\alpha}}, \qquad p>0,
\end{equation} 
which is sometimes referred to as the Kohlrausch-Williams-Watts function \cite{RSAnderssen04}. The divergence of all the moments of $g_{\alpha}(x)$ of positive integer order is related to a slow decay of $g_{\alpha}(x)$, ($g_{\alpha}(x) \to x^{-(1+\alpha)}$, as $x\to \infty$), usually termed as a "heavy-tail". Theoretical properties of $g_{\alpha}(x)$'s are discussed in \cite{APiryatinska05}, whereas the review of physical applications can be found in \cite{WAWoyczynski01} and \cite{IRoth11}. Many applications of $g_{\alpha}(x)$'s in financial mathematics can be found in \cite{RCont04}, whereas their use in various simulations of jump processes is exposed in \cite{JEFigueroa12}. The quest for exact and explicit forms of $g_{\alpha}(x)$ was accomplished only recently \cite{KAPenson10, KGorska11, ASaa11}. The universal formulas of Eqs.~(2), (3) and (4) in \cite{KAPenson10} give explicit forms of $g_{\alpha}(x)$ for fractional $\alpha = l/k$ with $l$ and $k$ arbitrary relatively prime integers, fulfilling the condition $l < k$. These results are satisfying as in the general case they furnish $g_{l/k}(x)$ as a sum of $k-1$ hypergeometric functions, which is a neat confirmation of the conjecture of Scher and Montroll \cite{HScher75}. As indicated in \cite{KAPenson10, KGorska11} the implementation of Meijer's G functions as well as of the hypergeometric functions in recent version of computer algebra systems \cite{Maple} greatly facilitates the practical use of $g_{l/k}(x)$ in applications.

The purpose of this work is to present an alternative approach to generate $g_{l/k}(x)$ by an iterative procedure based on the self-reproducing property of $g_{\alpha}(x)$, directly resulting from Eq.~(\ref{eq1}). We shall proceed in an increasing order of complexity treating first in Sect.~2 the low-order values of $\alpha$, ($\alpha = 1/2, 1/3$ and $2/3$), for which exact forms were obtained in \cite{KAPenson10, EBarkai01, EWMontroll84}, see also \cite{VVUchaikin99}. In Sect.~2 we also formulate the formalism for general $\alpha$ and develop the notion of L\'{e}vy transform (initiated in \cite{EBarkai01}), which will allow one the generation of general distributions $g_{l/k}(x)$ by simple integration. In Sect.~3 we present applications of our method. Sect.~4 is devoted to discussion and conclusions. The definition and notations of Meijer's G functions are given in the Appendix A. A~compact formulation of our results using the Efros theorem is presented in Appendix~B.


\section{Low-order $\alpha = l/k$ and reproducing properties of $g_{\alpha}(x)$} 

\noindent
\textbf{a)} The content of this subsection a) concerns the first genuine stable distribution $g_{\frac{1}{2}}(x)$ discovered by L\'{e}vy \cite{JPKahane95}, and therefore all the results enumerated here are known and are included only for sake of completeness. This historically first L\'{e}vy stable distribution is:
\begin{equation}\label{eq2}
g_{\frac{1}{2}}(x) = \frac{\exp\left(-\frac{1}{4\, x}\right)}{2\sqrt{\pi}\, x^{\frac{3}{2}}}, \quad x \geq 0,
\end{equation}
satisfying $\mathcal{L}\left[g_{\frac{1}{2}}(x); p\right] = e^{- \sqrt{p}}$, with $p > 0$. We can formally introduce an additional variable $t$ in the problem by writing 
\begin{equation}\label{eq3}
\frac{1}{t^{2}}\, g_{\frac{1}{2}}\left(\frac{x}{t^{2}}\right) = \frac{t}{2\sqrt{\pi}\, x^{\frac{3}{2}}}\, e^{-t^{2}/(4\,x)}.
\end{equation}
This last function is of particular interest for us as it intervenes directly in the following property of standard Laplace transform (see formula 1.1.1.26, p.~4, vol.~5 of \cite{APPrudnikov92}): if $\mathcal{L}\left[f(x); p\right] = F(p)$, then if we define $\tilde{f}_{\frac{1}{2}}(x)$ such that
\begin{equation}\label{eq4}
\tilde{f}_{\frac{1}{2}}(x) = \frac{1}{2\sqrt{\pi}\, x^{\frac{3}{2}}}\, \int_{0}^{\infty} t\, \exp\left(-\frac{t^{2}}{4\, x}\right)\, f(t)\, dt,
\end{equation}
then
\begin{equation}\label{eq5}
\mathcal{L}\left[\tilde{f}_{\frac{1}{2}}(x); p\right] = F(p^{\frac{1}{2}}).
\end{equation}
The proof of Eq.~(\ref{eq5}) follows:
\begin{eqnarray}\label{eq6}
\mathcal{L}\left[\tilde{f}_{\frac{1}{2}}(x); p\right] &=& \int_{0}^{\infty} e^{-p\, x}\, \left[\frac{1}{2\sqrt{\pi}\, x^{\frac{3}{2}}}\, \int_{0}^{\infty} t\, e^{-t^{2}/(4\, x)}\, f(t)\, dt \right]\, dx  \\[0.7\baselineskip] \label{eq7}
&=& \frac{1}{2\sqrt{\pi}}\, \int_{0}^{\infty} f(t)\,t\, \left[\int_{0}^{\infty} e^{-p x} e^{-t^{2}/(4 x)}\, \frac{d x}{x^{\frac{3}{2}}}\right]\, dt.
\end{eqnarray}
To evaluate the internal integration in Eq.~(\ref{eq7}) we use the formula 2.2.2.5, p.~31, vol.~4 of \cite{APPrudnikov92}, and Eq.~(\ref{eq7}) becomes 
\begin{equation}\label{eq8}
\frac{1}{2\sqrt{\pi}} \int_{0}^{\infty} t \, \left[\frac{2\sqrt{\pi}}{t}\, e^{-t\, \sqrt{p}}\right]\, f(t)\, dt = F(p^{\frac{1}{2}}),
\end{equation}
and therefore, from Eq.~(\ref{eq3})
\begin{equation}\label{eq9}
\mathcal{L}\left[\frac{1}{t^{2}}\, g_{\frac{1}{2}}\left(\frac{x}{t^{2}}\right); p\right] = \mathcal{L}\left[\frac{t\, \exp\left(-\frac{t^{2}}{4\, x}\right)}{2\sqrt{\pi}\, x^{\frac{3}{2}}}; p \right] = e^{-t\, \sqrt{p}}.
\end{equation}
We note that although all our PDF's are one-dimensional, $t^{-3}\, g_{\frac{1}{2}}(x/t^{2})$ is formally equal to the three-dimensional heat kernel, through appropriate renaming of the variables.

\noindent
\textbf{b)} As a next step in our development we consider the stable L\'{e}vy density for $\alpha = 1/3$ whose exact form is \cite{HScher75, VVUchaikin99}
\begin{equation}\label{eq10}
g_{\frac{1}{3}}(x) = \frac{1}{3\pi\, x^{\frac{3}{2}}}\, K_{\frac{1}{3}}\left(\frac{2}{3\sqrt{3\, x}}\right), \qquad x \geq 0,
\end{equation}
where $K_{\nu}(x)$ is the modified Bessel function of the second kind, which can also be expressed through the Airy function $Ai(y)$ giving 
\begin{equation}\label{eq10a}
g_{\frac{1}{3}}(x) = \frac{Ai\left((3 x)^{-\frac{1}{3}}\right)}{(3 x^4)^{\frac{1}{3}}}.
\end{equation}
We shall evaluate the Laplace transform of $g_{\frac{1}{3}}(x)$ using the formula 3.16.3.7, p.~355, vol.~4 of \cite{APPrudnikov92}, that we quote here \cite{Err}:
\begin{equation}\label{eq11}
\mathcal{L}\left[x^{-\frac{3}{2}}\, K_{\frac{1}{3}}\left(\frac{a}{\sqrt{x}}\right); p \right] = \frac{2 \pi}{a\, \sqrt{3}}\, \exp\left[-3 \left(\frac{a^{2}}{4}\right)^{\frac{1}{3}}\, p^{\frac{1}{3}}\right], \quad p > 0,
\end{equation}
which, upon choosing $3 (a^{2}/4)^{\frac{1}{3}} = t$, transforms into
\begin{equation}\label{eq12}
\mathcal{L}\left[\frac{1}{3 \pi}\, \left(\frac{t}{x}\right)^{\frac{3}{2}} K_{\frac{1}{3}}\left(\frac{2\, t^{\frac{3}{2}}}{3\sqrt{3\, x}}\right); p\right] = e^{-t\, p^{\frac{1}{3}}},
\end{equation}
or equivalently, can be rewritten as
\begin{equation}\label{eq13}
\mathcal{L}\left[\frac{1}{t^{3}}\, g_{\frac{1}{3}}\left(\frac{x}{t^{3}}\right); p\right] = e^{-t\, p^{\frac{1}{3}}}.
\end{equation}
In full analogy with Eqs.~(\ref{eq4}), (\ref{eq5}) and (\ref{eq6}) we conclude that if $\mathcal{L}[f(x); p] = F(p)$ then for $\tilde{f}_{\frac{1}{3}}(x)$ defined through
\begin{equation}\label{eq14}
\tilde{f}_{\frac{1}{3}}(x) = \frac{1}{3\pi\, x^{\frac{3}{2}}} \int_{0}^{\infty} t^{\frac{3}{2}}\, K_{\frac{1}{3}}\left(\frac{2\, t^{\frac{3}{2}}}{3\sqrt{3\, x}}\right)\, f(t)\, dt, \qquad x \geq 0,
\end{equation}
\begin{equation}\label{eq15}
\mathcal{L}[\tilde{f}_{\frac{1}{3}}(x); p] = F(p^{\frac{1}{3}}).
\end{equation}

\noindent
\textbf{c)} We continue our review of L\'{e}vy stable densities expressible with standard special functions with $g_{\frac{2}{3}}(x)$ which reads \cite{KAPenson10, EWMontroll84, VVUchaikin99, KGorska10-JMP}:
\begin{eqnarray}\label{eq16}
g_{\frac{2}{3}}(x) &=& \frac{2\sqrt{3}}{27 \pi x^{3}}\, e^{-\frac{2}{27\, x^2}} \left[K_{\frac{1}{3}}\left(\frac{2}{27\, x^2}\right) + K_{\frac{2}{3}}\left(\frac{2}{27\, x^2}\right)\right], \quad x\geq 0  \\[0.7\baselineskip]
&=& {\small \frac{\Gamma\left(\frac{2}{3}\right)}{\sqrt{3} \pi} x^{-\frac{5}{3}}\, _{1}F_{1}\left(\begin{array}{c} 5/6 \\ 2/3 \end{array} \Big | \footnotesize{\frac{-2^2}{3^3 x^2}}\right) + \frac{2/9}{\Gamma\left(\frac{2}{3}\right)} x^{-\frac{7}{3}}\, _{1}F_{1}\left(\begin{array}{c} 7/6 \\ 4/3 \end{array} \Big | \footnotesize{\frac{-2^2}{3^3 x^2}}\right)}, \label{eq17}
\end{eqnarray}
where in Eq.~(\ref{eq17}) $_{1}F_{1}(^{a}_{b}\big| z) = \,_{1}F_{1}(a; b; z) $ is Kummer's confluent hypergeometric function. In analogy with Eqs.~(\ref{eq3}) and (\ref{eq13}) we form $t^{-\frac{3}{2}} g_{\frac{2}{3}}(x/t^{\frac{3}{2}})$ in the version of Eq.~(\ref{eq17}):
\begin{equation}\label{eq18}
{\small \frac{1}{t^{\frac{3}{2}}}\, g_{\frac{2}{3}}\left(\frac{x}{t^{\frac{3}{2}}}\right) = \frac{\Gamma\left(\frac{2}{3}\right)}{\sqrt{3} \pi}\, \frac{t}{x^{\frac{5}{3}}}\, _{1}F_{1}\left(\begin{array}{c} 5/6 \\ 2/3 \end{array} \Big | \frac{-2^2 t^3}{3^3 x^2}\right)  + \frac{2/9}{\Gamma\left(\frac{2}{3}\right)} \frac{t^2}{x^{\frac{7}{3}}}\, _{1}F_{1}\left(\begin{array}{c} 7/6 \\ 4/3 \end{array} \Big | \frac{-2^2 t^3}{3^3 x^2}\right)}.
\end{equation}
To calculate the Laplace transform of Eq.~(\ref{eq18}) we apply the formula 3.35.1.16, p. 511, vol. 4 of \cite{APPrudnikov92} and, after simplifications, we arrive at the identity
\begin{equation}\label{eq19}
\mathcal{L}\left[\frac{1}{t^{\frac{3}{2}}}\, g_{\frac{2}{3}}\left(\frac{x}{t^{\frac{3}{2}}}\right); p\right] = e^{-t\, p^{\frac{2}{3}}}.
\end{equation}
This yields in turn, for $\mathcal{L}[f(x); p] = F(p)$, the pair of equations
\begin{equation}\label{eq20}
\tilde{f}_{\frac{2}{3}}(x) = \int_{0}^{\infty} \frac{1}{t^{\frac{3}{2}}}\, g_{\frac{2}{3}}\left(\frac{x}{t^{\frac{3}{2}}}\right)\, f(t)\, dt, \qquad x\geq 0
\end{equation}
and
\begin{equation}\label{eq21}
\mathcal{L}[\tilde{f}_{\frac{2}{3}}(x); p] = F(p^{\frac{2}{3}}),
\end{equation}
in complete analogy to the pairs of Eqs.~(\ref{eq4}), (\ref{eq9}) and Eqs.~(\ref{eq14}), (\ref{eq15}).

\noindent
\textbf{d)} The above pattern suggests the validity of the following general integral identities for $\mathcal{L}[f(x); p]~=~F(p)$: if
\begin{equation}\label{eq22}
\tilde{f}_{\alpha}(x) = \int_{0}^{\infty} \frac{1}{t^{\frac{1}{\alpha}}}g_{\alpha}\left(\frac{x}{t^{\frac{1}{\alpha}}}\right)\, f(t)\, dt, \qquad 0 < \alpha < 1, \qquad x\geq 0,
\end{equation}
then
\begin{equation}\label{eq23}
\mathcal{L}[\tilde{f}_{\alpha}(x); p] = F(p^{\alpha})\, .
\end{equation}
We are going to prove Eq.~(\ref{eq23}) using Eq.~(\ref{eq1}). To this end we calculate explicitly Eq.~(\ref{eq23}):
\begin{eqnarray}\label{eq24}
\mathcal{L}[\tilde{f}_{\alpha}(x); p] &=& \int_{0}^{\infty} e^{-p x} \left[\int_{0}^{\infty} \frac{1}{t^{\frac{1}{\alpha}}}\, g_{\alpha}\left(\frac{x}{t^{\frac{1}{\alpha}}}\right)\, f(t)\, dt\right]\, dx \\[0.7\baselineskip] \label{eq25}
&=& \int_{0}^{\infty} f(t)\, \left[\int_{0}^{\infty} e^{-p x} \frac{1}{t^{\frac{1}{\alpha}}}\, g_{\alpha}\left(\frac{x}{t^{\frac{1}{\alpha}}}\right)\, dx\right]\, dt \\[0.7\baselineskip] \label{eq26}
&=& \int_{0}^{\infty} f(t)\, \left[\int_{0}^{\infty} e^{-p t^{\frac{1}{\alpha}} y} g_{\alpha}(y)\, dy\right]\, dt \\[0.7\baselineskip] \label{eq27}
&=& \int_{0}^{\infty} f(t)\, e^{- p^{\alpha} t}\, dt = F(p^{\alpha}) \, .
\end{eqnarray}
In Eq.~(\ref{eq25}) we have used a simple change of variable, whereas in Eq.~(\ref{eq26}) we have applied Eq.~(\ref{eq1}).

Eqs.~(\ref{eq22}) and (\ref{eq23}) acquire additional importance when for $f(t)$ we choose another L\'{e}vy stable density, say $g_{\beta}(x)$ with arbitrary $\beta$, such that $0 < \beta < 1$. Then
\begin{equation}\label{eq29}
\int_{0}^{\infty} \frac{1}{t^{\frac{1}{\alpha}}} g_{\alpha}\left(\frac{x}{t^{\frac{1}{\alpha}}}\right)\,g_{\beta}(t)\, dt = \int_{0}^{\infty} \frac{1}{t^{\frac{1}{\beta}}} g_{\beta}\left(\frac{x}{t^{\frac{1}{\beta}}}\right)\,g_{\alpha}(t)\, dt, \quad x\geq 0,
\end{equation}  
which defines the following transitive property of L\'{e}vy laws:
\begin{equation}\label{eq30}
g_{\alpha\, \beta}(x) = \int_{0}^{\infty} \frac{1}{t^{\frac{1}{\alpha}}} g_{\alpha}\left(\frac{x}{t^{\frac{1}{\alpha}}}\right)\,g_{\beta}(t)\, dt = \int_{0}^{\infty} \frac{1}{t^{\frac{1}{\beta}}} g_{\beta}\left(\frac{x}{t^{\frac{1}{\beta}}}\right)\,g_{\alpha}(t)\, dt,
\end{equation}
for $0 < \alpha, \beta < 1$. Eqs.~(\ref{eq23}) and (\ref{eq30}) constitute the key results of the present investigation. The following remarks are in order.
\begin{itemize}
\item[i)] The formulas Eqs.~(\ref{eq4}) and (\ref{eq5}) deserve to be better known. They appear (along with their several extensions) in various works of Russian school \cite{APPrudnikov92, VSMartynenko68, VADitkin65}, but are conspicuously absent from other monographs and handbooks \cite{LDebnath06, INSneddon72}.

\item[ii)] the formulas Eqs.~(\ref{eq15}), (\ref{eq19}) and more generally, Eqs.~(\ref{eq22}) and (\ref{eq23}) should complement the formula of Eq.~(\ref{eq5}) in the lists of properties of Laplace transform. In subsequent Sections we obtain many other formulas of this type, by specifying other values of $\alpha$ and $\beta$.

\item[iii)] Eq.~(\ref{eq30}) should be viewed as a tool to generate the L\'{e}vy stable PDF's, starting with those with arbitrary $\alpha$ and $\beta$ and yielding one with $\alpha\,\beta$, via integration.
\end{itemize}

The content of Eqs.~(\ref{eq22}) and (\ref{eq23}) can be seen as an integral transform with the positive kernel $\kappa_{\alpha}(t, x) = t^{-\frac{1}{\alpha}} g_{\alpha}(x/t^{\frac{1}{\alpha}})$ with $x, t > 0$, which we define as the L\'{e}vy2 transform of index $\alpha$, and is denoted by $L_{\alpha}^{(2)}$:
\begin{equation}\label{eq32}
L_{\alpha}^{(2)}[f(t); x] = \int_{0}^{\infty} \kappa_{\alpha}(t, x)\, f(t)\, dt = \tilde{f}_{\alpha}(x).
\end{equation}
According to Eq.~(\ref{eq30}) the action of $L_{\alpha}^{(2)}$ on L\'{e}vy's PDFs satisfy
\begin{equation}\label{eq33}
L_{\alpha}^{(2)}[g_{\beta}(t); x] = L_{\beta}^{(2)}[g_{\alpha}(t); x] = g_{\alpha \beta}(x),
\end{equation}
which well illustrates the reproducing property of L\'{e}vy densities under the L\'{e}vy2 transforms. It can be seen from Eqs.~(\ref{eq32}) and (\ref{eq33}) that for $0 < \alpha < 1$ the inverse of L\'{e}vy2 transform cannot be meaningfully defined. As a consequence, Eq.~(\ref{eq33}) reflects the semigroup property of $L_{\alpha}^{(2)}$ in accordance with the general theory \cite{WFeller71}.

Alternative way to derive the results of this section is to use the Efros theorem of theory of Laplace transform, see Appendix B.

\section{Some applications of the method}

\noindent
\textbf{a)} We shall exemplify now how the method works by choosing in Eq.~(\ref{eq33}) $\alpha = \beta = 1/2$ which accordingly should furnish $g_{\frac{1}{4}}(x)$. We use Eqs.~(\ref{eq2}) and (\ref{eq3}) and obtain
\begin{equation}\label{eq35}
g_{\frac{1}{4}}(x) = \frac{1}{4 \pi\, x^{\frac{3}{2}}} \int_{0}^{\infty} t^{-\frac{1}{2}} \exp\left[-\left(\frac{t^2}{4 x} + \frac{1}{4 t}\right)\right]\, dt.
\end{equation}
The integral in Eq.~(\ref{eq35}) is not elementary, but with a new variable $y = t^2$ it becomes a particular case of the formula 2.3.2.14, p.~322, vol.~1 of \cite{APPrudnikov98} (the alternative way is~to employ the formula 2.2.2.7, p.~32, vol.~4 of \cite{APPrudnikov92}), which involves a finite sum of hypergeometric functions. The final answer, after the simplifications in hypergeometric functions $\,_{p}F_{q}\left(^{(\alpha_{p})}_{(\beta_{q})} \Big| z\right) = \,_{p}F_{q}\left(\alpha_{1}, \ldots, \alpha_{p};\, \beta_{1}, \ldots, \beta_{q};\, z\right)$, is:
\begin{eqnarray}\label{eq36}
{\small g_{\frac{1}{4}}(x)} &=& {\small \frac{\Gamma\left(\frac{3}{4}\right) }{2^{\frac{7}{2}} \pi\, x^{\frac{7}{4}}}  \,_{0}F_{2}\left(\begin{array}{c} \_\_\_ \\ \frac{5}{4}, \frac{3}{2} \end{array} \Big| {\footnotesize\frac{-1}{2^8 x}} \right) - \frac{1}{4 \sqrt{\pi}\, x^{\frac{3}{2}}} \,_{0}F_{2}\left(\begin{array}{c} \_\_\_ \\ \frac{3}{4}, \frac{5}{4} \end{array} \Big| {\footnotesize \frac{-1}{2^8 x}} \right)}  \nonumber\\[0.7\baselineskip]
&+& {\small \frac{1}{4 \Gamma\left(\frac{3}{4}\right)\, x^{\frac{5}{4}}} \,_{0}F_{2}\left(\begin{array}{c} \_\_\_ \\ \frac{1}{2}, \frac{3}{4} \end{array} \Big| {\footnotesize \frac{-1}{2^8 x}} \right),} 
\end{eqnarray}
which is precisely the result obtained in \cite{KAPenson10} and \cite{EBarkai01}. We stress that here it was calculated with one definite integration using the tables \cite{APPrudnikov98}.

\noindent
\textbf{b)} Next example in this spirit will involve the indices $\alpha = 1/2$ and $\beta = 1/3$, i. e. we shall explicitly perform the L\'{e}vy2 transforms of type
\begin{equation}\label{eq37}
L_{\frac{1}{2}}^{(2)}[g_{\frac{1}{3}}(t); x] = L_{\frac{1}{3}}^{(2)}[g_{\frac{1}{2}}(t); x] = g_{\frac{1}{6}}(x),
\end{equation}
with Eqs.~(\ref{eq3}) and (\ref{eq10}). The result reads ($b \equiv \frac{2}{3\sqrt{3}}$)
\begin{equation}\label{eq38}
g_{\frac{1}{6}}(x) = \frac{1}{2 \sqrt{\pi}\, x^{\frac{3}{2}}}\, \int_{0}^{\infty} t\, e^{-t^{2}/(4 x)} \left[\frac{1}{3\pi t^{\frac{3}{2}}}\, K_{\frac{1}{3}}\left(\frac{b}{\sqrt{t}}\right)\right]\, dt,
\end{equation}
which, with a change of variable, transforms into
\begin{equation}\label{eq39}
g_{\frac{1}{6}}(x) = \frac{1}{12 (\pi\,x)^{\frac{3}{2}}}\, \int_{0}^{\infty} e^{-y/(4 x)} \left[y^{-\frac{3}{4}}\, K_{\frac{1}{3}}\left(\frac{b}{y^{\frac{1}{4}}}\right)\right]\, dy.
\end{equation}
The integral in Eq.~(\ref{eq39}) can be perceived as the Laplace transform of $y^{-\frac{3}{4}} K_{\frac{1}{3}}(b\, y^{-1/4})$ for which an exact formula in terms of a specific Meijer's G function exists, compare Eq.~3.16.3.9, p. 355, vol. 4 of \cite{APPrudnikov92}, with the specializations: $a = b = \frac{2}{3\sqrt{3}}$, $l = 1$, $k = 2$, $\mu = -3/4$, $\nu = 1/3$ and $p = 1/(4x)$. The final result is proportional to the Meijer's G function of type {\small$G^{\,0, 5}_{\,5, 0}\left({\footnotesize 6^6\, x \,\Big| \begin{array}{c} \frac{5}{12}, \frac{7}{12}, \frac{3}{4}, \frac{11}{12}, \frac{13}{12}\\ \_\_\_\_\_\_ \end{array}} \right)$}, see Appendix.  By lumping together all the constants we obtain:
\begin{eqnarray}\label{eq40}
{\small g_{\frac{1}{6}}(x)} &=& {\small \frac{\sqrt{2}}{48 \pi^{5/2} x^{5/4}}\, G^{\,0, 5}_{\,5, 0}\left({\footnotesize 6^6\, x} \,\Big| \begin{array}{c} \frac{5}{12}, \frac{7}{12}, \frac{3}{4}, \frac{11}{12}, \frac{13}{12}\\ \_\_\_\_\_\_ \end{array} \right)} \\[0.7\baselineskip]
&=& {\small \frac{2^{-\frac{1}{3}} 3^{-\frac{3}{2}} \sqrt{\pi}}{\left[\Gamma\left(\frac{2}{3}\right)\right]^{2} x^{\frac{7}{6}}} \,_{0}F_{4}\left(\begin{array}{c} \_\_\_\_ \\ \frac{1}{3}, \frac{1}{2}, \frac{2}{3}, \frac{5}{6}\end{array} \Big| \frac{-1}{6^6 x}\right) - \frac{1}{6\Gamma\left(\frac{2}{3}\right)\, x^{\frac{4}{3}}} \,_{0}F_{4}\left(\begin{array}{c} \_\_\_\_ \\ \frac{1}{2}, \frac{2}{3}, \frac{5}{6}, \frac{7}{6}\end{array} \Big| \frac{-1}{6^6 x}\right)} \nonumber\\[0.7\baselineskip]
&+& {\small \frac{(12)^{-1}}{\sqrt{\pi}\, x^{\frac{3}{2}}} \,_{0}F_{4}\left(\begin{array}{c} \_\_\_\_ \\ \frac{2}{3}, \frac{5}{6}, \frac{7}{6}, \frac{4}{3}\end{array} \Big| \frac{-1}{6^6 x}\right) - \frac{\sqrt{3}\, \Gamma\left(\frac{2}{3}\right)}{72 \pi x^{\frac{5}{3}}} \,_{0}F_{4}\left(\begin{array}{c} \_\_\_\_ \\ \frac{5}{6}, \frac{7}{6}, \frac{4}{3}, \frac{3}{2}\end{array} \Big| \frac{-1}{6^6 x}\right)} \nonumber\\[0.7\baselineskip]
&+& \frac{3^{-\frac{3}{2}}\left[\Gamma\left(\frac{2}{3}\right)\right]^{2}}{2^{\frac{17}{3}}\pi^{\frac{3}{2}} x^{\frac{11}{6}}} \,_{0}F_{4}\left(\begin{array}{c} \_\_\_\_ \\ \frac{7}{6}, \frac{4}{3}, \frac{3}{2}, \frac{5}{3}\end{array} \Big| \frac{-1}{6^6 x}\right) \label{eq41}.
\end{eqnarray}
The Eq.~(\ref{eq41}) was obtained using the formulas 16.17.2 and 16.17.3 of \cite{NIST}. The result of Eq.~(\ref{eq41}) is a special case of the solutions in \cite{KAPenson10}. The explicit form of $g_{\frac{1}{6}}(x)$ in Eq.~(\ref{eq41}) appears to be written down for the first time here.

\noindent
\textbf{c)} The third example in this section will be of more academic value as it will concern three PDF's known from the previous sections: $g_{\frac{1}{2}}(x)$, $g_{\frac{2}{3}}(x)$ and $g_{\frac{1}{3}}(x)$. According to Eqs.~(\ref{eq30}) and (\ref{eq33}) they are related through
\begin{equation}\label{eq42}
L_{\frac{1}{2}}^{(2)}[g_{\frac{2}{3}}(t); x] = L_{\frac{2}{3}}^{(2)}[g_{\frac{1}{2}}(t); x] = g_{\frac{1}{3}}(x). 
\end{equation}
Indeed, the first integral transform of Eq.~(\ref{eq42}) is equal to
\begin{eqnarray}\label{eq43}
&\int_{0}^{\infty} \frac{1}{t^{\frac{3}{2}}}& g_{\frac{2}{3}}\left(\frac{x}{t^{\frac{3}{2}}}\right) g_{\frac{1}{2}}(t)\, dt = \frac{\Gamma\left(\frac{2}{3}\right)}{2\sqrt{3}\,\pi\, x^{\frac{5}{3}}} \int_{0}^{\infty} t^{-\frac{1}{2}}\, e^{-\frac{1}{4 x}}\,_{1}F_{1}\left(\begin{array}{c} 5/6 \\ 2/3\end{array} \Big| \frac{-2^2 t^3}{3^3 x^2} \right)\, dt \nonumber\\[0.7\baselineskip]
& & + \frac{1}{9\sqrt{\pi}\Gamma\left(\frac{2}{3}\right)\, x^{7/3}} \int_{0}^{\infty} t^{\frac{1}{2}}\, e^{-\frac{1}{4 x}}\,_{1}F_{1}\left(\begin{array}{c} 7/6 \\ 4/3\end{array} \Big| \frac{-2^2 t^3}{3^3 x^2} \right)\, dt, 
\end{eqnarray}
and, with the help of the formula 3.38.1.30, p. 553, vol. 4 of \cite{APPrudnikov92}, it can be written down as a special case of the Meijer's G function, very much in the spirit of considerations leading to Eq.~(\ref{eq40}), however see \cite{Err1}. (Alternatively the simplified formula 3.35.1.16, p.~511, vol.~4 of \cite{APPrudnikov92} can also be used.). We will skip further details of this evaluation and conclude that the result is equal to $g_{\frac{1}{3}}(x)$ from Eq.~(\ref{eq10}), as it should be.

\noindent
\textbf{d)} In the following examples we shall not use the reproducing property of Eq.~(\ref{eq30}). We shall instead concentrate on Eqs.~(\ref{eq22}) and (\ref{eq23}) conceived as an operational property of the Laplace transform. To this end we, quite arbitrarily, choose the modified Bessel function $K_{0}(x)$ as $f(x)$ in Eq.~(\ref{eq22}) and we read off its Laplace transform from the Eq.~3.16.1.2, p. 349, vol. 4 of \cite{APPrudnikov92}, which is 
\begin{equation}\label{e1}  
\mathcal{L}\left[K_{0}(x);\, p\right] = \frac{\arccos(p)}{\sqrt{1 - p^{2}}}, \quad p > 0.
\end{equation}
If we set $g_{\frac{1}{2}}(z)$ in Eq.~(\ref{eq22}) then the corresponding transformed function is
\begin{eqnarray}\label{e2}
\tilde{K}_{0, \frac{1}{2}}(x) &=& \frac{e^{x} \Gamma(0, x)}{2\, \sqrt{\pi x}} \\[0.6\baselineskip]
&=& -\frac{e^{x} Ei(-x)}{2\sqrt{\pi x}}. \label{e3}
\end{eqnarray} 
In Eq.~(\ref{e2}) we have used the formula 2.16.8.5, p. 352, vol. 2 of \cite{APPrudnikov98}, with $\Gamma(0, x)~=~-Ei(-x)$, see p. 726, vol. 2 of \cite{APPrudnikov98}. In Eqs.~(\ref{e2}) and (\ref{e3}), $\Gamma(\nu, z)$ and $Ei(z)$ are incomplete gamma function and the exponential integral, respectively. The Laplace transform of the function in Eq.~(\ref{e3}) can be evaluated to be
\begin{equation}\label{e4}
\mathcal{L}\left[\tilde{K}_{0, \frac{1}{2}}(x); p\right] = \frac{\mathrm{arcsinh}(\sqrt{p-1})}{\sqrt{p-1}}, \quad p > 0.
\end{equation}
Let us briefly sketch how Eq.~(\ref{e4}) comes about. \\
We write out explicitly the Laplace transform in Eq.~(\ref{eq4}) as
\begin{eqnarray}\label{e5}
\mathcal{L}\left[\tilde{K}_{0, \frac{1}{2}}(x), p\right] &=& - \int_{0}^{\infty} \frac{e^{-(p-1)} Ei(-x)}{2 \sqrt{\pi x}}\, dx \\[0.6\baselineskip]
&=& \frac{1}{\sqrt{p}} \,_{2}F_{1}\left(\begin{array}{c} \frac{1}{2}, 1 \\ \frac{3}{2}\end{array} \Big| \frac{p-1}{p} \right) \label{e6} \\[0.6\baselineskip]
&=& \frac{\mathrm{arcsinh}\left(\sqrt{p-1}\right)}{\sqrt{p-1}}, \quad p > 0. \label{e7}
\end{eqnarray}   
In Eq.~(\ref{e6}) we have used the Eq.~3.4.1.3, p. 135, vol. 4 of \cite{APPrudnikov92}, whereas Eq.~(\ref{e7}) results from Eq.~7.3.2.83, p. 473, vol. 3 of \cite{APPrudnikov98}. Since for $p > 0$, $\arccos(\sqrt{p})/\sqrt{1-p}~=~\mathrm{arcsinh}(\sqrt{p-1})/\sqrt{p-1}$, the above calculations evidently confirm the validity of Eq.~(\ref{eq23}). \\
We go now a step further: while still keeping $f(x) = K_{0}(x)$ in Eq.~(\ref{eq23}), we use now $g_{\frac{1}{3}}(z)$, see Eq.~(\ref{eq10}), to perform the integration in Eq.~(\ref{eq22}). The corresponding transformed function $\tilde{K}_{0, \frac{1}{3}}(x)$ reads \cite{Maple}
\begin{equation}\label{e8}
\tilde{K}_{0, \frac{1}{3}}(x) = \frac{\sqrt{3/2}}{8\, \pi^{4} x^{3/2}}\, G^{\, 6, 4}_{\, 4, 6}\left(\frac{x^{2}}{4} \Big| \begin{array}{c}\frac{5}{12}, \frac{7}{12}, \frac{11}{12}, \frac{13}{12} \\ \frac{5}{12}, \frac{5}{12}, \frac{3}{4}, \frac{3}{4}, \frac{13}{12}, \frac{13}{12} \end{array}\right), \quad x > 0,
\end{equation}  
whose Laplace transform can be calculated via 3.40.1.1, vol. 4 of \cite{APPrudnikov92}. This lengthy expression involves a combination of four different hypergeometric functions of argument $p^{2}$. We shall not reproduce it here. We emphasize however that we have verified numerically the relation
\begin{equation}\label{e9}
\mathcal{L}\left[\tilde{K}_{0, \frac{1}{3}}(x); p\right] = \frac{\arccos\left(p^{1/3}\right)}{\sqrt{1 - p^{2/3}}}, \quad p > 0,
\end{equation}
thereby establishing an algebraic identity stating that the aforementioned combination of hypergeometric functions is equal to the r.h.s. of Eq.~(\ref{e9}).\\
It is clear that these procedures can be carried out by choosing functions other than $K_{0}(x)$ in Eq.~(\ref{eq22}). This circumstance paves the way for a scheme of generation of algebraic identities involving special functions using Eqs.~(\ref{eq22}) and (\ref{eq23}). This will be a subject of future publication.

\section{Discussion and Conclusions}

Our kernel $\kappa_{\alpha}(t, x)$ which defines the L\'{e}vy2 transform of index $\alpha$ through Eq.~(\ref{eq32}) above, can be related to the function $n(s, \tau)$ introduced by E.~Barkai in \cite{EBarkai01} as follows
\begin{equation}\label{eq44}
n(s, \tau) = \frac{1}{\alpha\, s}\, \frac{\tau}{s^{\frac{1}{\alpha}}}\, g_{\alpha}\left(\frac{\tau}{s^{\frac{1}{\alpha}}}\right) = \frac{1}{\alpha\, s}\, \kappa_{\alpha}(\tau, s),
\end{equation}
where attention should be given to the order of variables in Eq.~(\ref{eq44}). This function serves as a kernel in another integral transform \cite{EBarkai01} which links a solution of ordinary Fokker-Planck equation $P_{1}(x, \tau)$, satisfying, for given functions $A(x)$ and $B(x)$,
\begin{equation}\label{eq45}
\frac{\partial}{\partial\tau} P_{1}(x, \tau) = - \frac{\partial}{\partial x}\left[A(x) P_{1}(x, \tau)\right] + \frac{\partial^{2}}{\partial x^{2}}\left[B(x) P_{1}(x, \tau)\right],
\end{equation}
with $P_{\alpha}(x, \tau)$ satisfying the \textit{fractional} Fokker-Planck (FFP) equation,
\begin{equation}\label{eq46}
\frac{\partial^{\alpha}}{\partial\tau^{\alpha}} P_{\alpha}(x, \tau) = - \frac{\partial}{\partial x}\left[A(x) P_{\alpha}(x, \tau)\right] + \frac{\partial^{2}}{\partial x^{2}}\left[B(x) P_{\alpha}(x, \tau)\right],
\end{equation}
where $\frac{\partial^{\alpha}}{\partial \tau^{\alpha}}$ is appropriately defined fractional derivative. The relation is 
\begin{equation}\label{eq47}
P_{\alpha}(x, \tau) = \int_{0}^{\infty} n(s, \tau)\, P_{1}(x, s)\, ds,
\end{equation}
and is called in \cite{EBarkai01} the inverse L\'{e}vy transform. The functional relation of this type appears also in Sokolov's work \cite{IMSokolov01}, section III, Eqs.~(15) and (16) of \cite{IMSokolov01}. If we provisionally reserve the notation $\left(L^{(1)}_{\alpha}\right)^{-1}$ to the transform in Eq.~(\ref{eq47}), then it can be rewritten as
\begin{equation}\label{eq48}
\left(L^{(1)}_{\alpha}\right)^{-1} \left[P_{1}(x, s); \tau\right] = P_{\alpha}(x, \tau).
\end{equation}
The symbol $L^{(1)}_{\alpha}$ should be clearly distinguished from the L\'{e}vy2 transform appearing in Eqs.~(\ref{eq32}) and (\ref{eq37}), denoted by $L^{(2)}_{\alpha}$. From Eq.~(\ref{eq44}) the relation between those two transforms is 
\begin{equation}\label{eq49}
L^{(2)}_{\alpha}\left[\frac{1}{\alpha\, s} P_{1}(x, s); \tau \right] = \left(L^{(1)}_{\alpha}\right)^{-1}\left[P_{1}(x, s), \tau\right],
\end{equation} 
where it should be stressed that in both sides of Eq.~(\ref{eq49}) the integration variable is $s$. Here for the general $\alpha$ the direct L\'{e}vy transform $L_{\alpha}^{(1)}$ cannot be constructed, compare remarks after Eq.~(\ref{eq33}).

In both Refs.~\cite{EBarkai01} and \cite{IMSokolov01} the emphasis is on generation of solutions of FFP equations depending on the choice of $A(x)$, the drift function, and $B(x)$, the diffusion function, whereas our method here focuses on generation of explicit forms of $g_{\alpha}(x)$. In the separate work we have undertaken the analysis of FFP equation using the explicit form of $g_{\frac{l}{k}}(x)$, see \cite{KGorska11-1}. Further considerations on applications of relations of type Eq.~(\ref{eq47}) can be found in \cite{DBrockmann02} and \cite{IMSokolov04}. Extensions of the current framework to include two-sided L\'{e}vy distributions \cite{KGorska11, ASaa11} as well as log-L\'{e}vy distributions \cite{IEliazar11} are under study. Finally, note that newer versions of Mathematica$^{\tiny{\textregistered}}$ furnish numerical forms of $g_{\alpha}(x)$. Similarly, Matlab$^{\tiny{\textregistered}}$ can be also used for that purpose.


\section{Acknowledgements}

We thank Ch.~Vignat for informing us about remarks in \cite{WFeller71}, p. 176 concerning existence of relations obtained in Eq.~(\ref{eq30}).

The authors acknowledge support from Agence Nationale de la Recherche (Paris, France) under Program No. ANR-08-BLAN-0243-2 and from PAN/CNRS Project PICS No.~4339. K.~G\'{o}rska acknowledges support from Funda\c{c}\~{a}o de Amparo \'{a} Pesquisa do Estado de S\~{a}o Paulo (FAPESP, Brasil) under Program No.~2010/15698-5.


\appendix
\section{Definition and notations of Meijer's G functions}

The Meijer's G function is defined as an inverse Mellin transform denoted by $\mathcal{M}^{-1}$, see vol. 3 of \cite{APPrudnikov98}:
\begin{eqnarray}\label{A1}
{\small G^{\,m, n}_{p, q}}{\small \left( z \Big\vert^{\Large{\,(\alpha_{p})}}_{\Large{\, (\beta_{q})}}\right) = \mathcal{M}^{-1} \left[\frac{\prod_{j=1}^{m}{\footnotesize \Gamma(\beta_{j}+s)} \prod_{j=1}^{n} {\footnotesize \Gamma(1-\alpha_{j}-s)}}{\prod_{j=m+1}^{q}{\footnotesize \Gamma(1-\beta_{j}-s)} \prod_{j=n+1}^{p}{\footnotesize \Gamma(\alpha_{j}+s)}} ; z\right]} \\[0.7\baselineskip]\label{A2}
= {\small G([[\alpha_{1},\ldots, \alpha_{n}], [\alpha_{n+1},\ldots, \alpha_{p}]], [[\beta_{1}, \ldots, \beta_{m}], [\beta_{m+1}, \ldots, \beta_{q}]], z),}
\end{eqnarray}
where in Eq.~(\ref{A1}) empty products are taken to be equal to one. In Eqs.~(\ref{A1}) and (\ref{A2}) the parameters are subject of conditions:
\begin{eqnarray}\label{A3}
&& z \neq 0,\,\,  0\leq m \leq q,\,\,\, 0 \leq n \leq p; \nonumber \\[0.6\baselineskip]\nonumber &&\alpha_{j}\in\mathbb{C}, \,\, j=1, \ldots, p;\,\,\, \beta_{j}\in\mathbb{C}, \,\, j=1, \ldots, q; 
\\[0.6\baselineskip] 
&& (\alpha_{p}) = \alpha_{1}, \alpha_{2}, \ldots, \alpha_{p}; \,\,\, (\beta_{q}) = \beta_{1}, \beta_{2}, \ldots, \beta_{q}.
\end{eqnarray}
For a full description of integration contours in Eq.~(\ref{A1}), general properties and special cases of the $G$ functions, see vol. 3 of \cite{APPrudnikov98}. In Eq.~(\ref{A2}) we present a transparent notation inspired by computer algebra systems \cite{Maple}.

In this context we can give more details on how Eq.~(\ref{eq40}) comes about. The rewriting again of Eq.~3.16.3.9, p. 355, vol. 4 of \cite{APPrudnikov92} with the set of parameters given above, gives the following expresion for $g_{\frac{1}{6}}(x)$:
\begin{equation}\label{A4}
g_{\frac{1}{6}}(x) \sim G^{\,0, 5}_{\,5, 0}\left(6^6 x\Big|\begin{array}{c} \frac{5}{12}, \frac{7}{12}, \frac{3}{4}, \frac{11}{12}, \frac{13}{12} \\ \_\_\_\_\_ \end{array}\right) .
\end{equation}
According to (\ref{A1}), in $G^{\,0, 5}_{\,5, 0}$ the parameters are: $m=0$, $n=5$, $p=5$, and $q=0$. If we apply this to the notation specified in the function $G^{\,0, 5}_{\,5, 0}$ in (\ref{A4}), it becomes:
\begin{eqnarray}\label{A5}
&G^{\,0, 5}_{\,5, 0}&\left(6^6 x\Big|\begin{array}{c} \frac{5}{12}, \frac{7}{12}, \frac{3}{4}, \frac{11}{12}, \frac{13}{12} \\ \_\_\_\_\_ \end{array}\right) \nonumber\\[0.6\baselineskip]
&&\qquad\qquad\equiv  G\left(\left[\left[{\small \frac{5}{12}, \frac{7}{12}, \frac{3}{4}, \frac{11}{12}, \frac{13}{12} }\right], \left[\,\atop\,\right]\right], [[\,\,\,], [\,\,\,]],6^6 x\right),
\end{eqnarray}
which is a required format of computer algebra systems \cite{Maple}. 

All the Meijer's G functions encountered in course of integrations via L\'{e}vy2 transform of Eq.~(\ref{eq32}) were converted to forms of type (\ref{A5}) and consequently checked numerically. 


\section{The Efros Theorem}

We quote here the A.~M.~Efros theorem (1935) which is the generalisation of the convolution theorem for the Laplace transform. See Refs.~\cite{VSMartynenko68, MLavrentiev77, UGraf04} for proof and many applications.

\noindent
\textit{Theorem:} If $G(p)$ and $q(p)$ are analytic functions, $\mathcal{L}\left[f(x); p\right] = F(p)$ and 
\begin{equation}\label{B1}
\mathcal{L}\left[g(x, t); p\right] = \int_{0}^{\infty} g(x, t)\, e^{- p x}dx = G(p)\, e^{-t q(p)},
\end{equation}
then 
\begin{equation}\label{B2}
G(p)\, F\left(q(p)\right) = \int_{0}^{\infty} \left[\int_{0}^{\infty} f(t)\, g(x, t)\, dt\right]\, e^{-p x} dx.
\end{equation}
We shall demonstrate now that Eq.~(\ref{eq30}) can be recast as a special case of the Efros theorem, Eq.~(\ref{B2}).

The use of the result (\ref{B2}) in the context of L\'{e}vy stable laws entails the choice $f(t) = g_{\beta}(t)$, $0 < \beta < 1$, implying $F(p) = e^{-p^{\beta}}$ with Eq.~(\ref{eq1}). Furthermore, we set $G(p) = 1$, $q(p) = p^{\alpha}$ and
\begin{equation}\label{eqB3}
g(x, t) = \frac{1}{t^{1/\alpha}}\, g_{\alpha}\left(\frac{x}{t^{1/\alpha}}\right), \qquad 0 < \alpha < 1.
\end{equation}
From the Efros theorem (\ref{B2}) we obtain
\begin{equation}\label{B4}
G(p)\, F\left(q(p)\right) = e^{-p^{\alpha \beta}} = \int_{0}^{\infty} \left[\int_{0}^{\infty} g_{\beta}(t)\, \frac{1}{t^{1/\alpha}}\, g_{\alpha}\left(\frac{x}{t^{1/\alpha}}\right)\, dt\right]\, e^{-p x} dx. 
\end{equation}
On the other hand, Eq.~(\ref{eq1}) implies
\begin{equation}\label{B5}
e^{-p^{\alpha \beta}} = \int_{0}^{\infty} g_{\alpha \beta}(x)\, e^{- p x} dx,
\end{equation}
which immediately yields
\begin{equation}\label{B6}
g_{\alpha \beta}(x) = \int_{0}^{\infty} g_{\beta}(t)\, \frac{1}{t^{1/\alpha}}\, g_{\alpha}\left(\frac{x}{t^{1/\alpha}}\right)\, dt,
\end{equation}
which is our Eq.~(\ref{eq30}).



\begin{thebibliography}{39}
\bibitem{Maple} We have made extensive use of Maple$^{\tiny{\textregistered}}$ and Mathematica$^{\tiny{\textregistered}}$ in this work.
\bibitem{Err} We have corrected a misprint in this formula: the correct prefactor in the numerator of r.h.s. of Eq.~(12) is $2 \pi$ and not $4 \pi$ as indicated in \cite{APPrudnikov92}.
\bibitem{Err1} Two misprints in this last formula should be corrected: in the prefactor of the r.h.s. $k^{\nu}$ should be replaced by $k^{\mu}$ and in the parameter list of the G function the (meaningless) symbol $\Delta(1 - (b_{n}))$ should be replaced by $\Delta(k, 1 - (b_{n}))$, where $\Delta\left(k, (a_{p})\right) = \frac{(a_{p})}{k}, \frac{(a_{p}) + 1}{k}, \ldots, \frac{(a_{p}) + k - 1}{k}$, see p. 798 of vol. 3 of \cite{APPrudnikov98}.
\end{thebibliography}
\end{document}